\documentclass[%
reprint,
amsmath,
amssymb,
amsfonts,
aps,
prb,
floatfix,
]{revtex4-2}
\usepackage{graphicx}
\usepackage{subfigure}  
\usepackage{dcolumn}
\usepackage{bm}
\usepackage{physics}
\usepackage{multirow}

\begin{document}


\title{Optical anisotropies of asymmetric double GaAs (001) Quantum Wells }

\author{O. Ruiz-Cigarrillo}
\email{oscarruiz@cactus.iico.uaslp.mx}
\author{L.F. Lastras-Mart\'inez}
\email{lflm@cactus.iico.uaslp.mx}
\author{E.A. Cerda-M\'endez}
\author{G. Flores-Rangel}
\author{C.A. Bravo-Velazquez}
\author{R.E. Balderas-Navarro}
\author{A. Lastras-Mart\'inez}
\affiliation{Instituto de Investigaci\'on en Comunicaci\'on \'Optica, Universidad Aut\'onoma de San Luis Potos\'i\\ Alvaro Obreg\'on 64, 78000 San Luis Potos\'i, S.L.P., M\'exico
}%
\author{N.A. Ulloa-Castillo}
\affiliation{Tecnologico de Monterrey, School of Engineering and Science, Av. Eugenio Garza Sada Sur 2501, Monterrey, Nuevo Le\'on 64849, M\'exico}
\author{K. Biermann}
\author{P.V. Santos}
\email{santos@pdi-berlin.de}
\affiliation{
Paul-Drude-Institut f\"ur Festk\"orperelektronik, Leibniz-Institut im Forschungsverbund Berlin e.V, Germany
}%

\date{\today}

\begin{abstract}
In the present work, we were able to identify and characterize a new source of in-plane optical anisotropies (IOAs) occurring in asymmetric DQWs; namely a reduction of the symmetry from $D_{2d}$ to $C_{2v}$ as imposed by asymmetry along the growth direction. We report  on reflectance anisotropy spectroscopy (RAS) of  double GaAs quantum wells (DQWs)  structures coupled by a thin ($<2$ nm) tunneling  barrier. Two groups of DQWs systems were studied: one where both QWs have the same thickness (symmetric DQW) and another one where they have different thicknesses (asymmetric DQW). RAS measures the IOAs arising from the intermixing of the heavy- and light- holes in the valence band when the symmetry of the DQW system is lowered from $D_{2d}$ to $C_{2v}$. If the DQW is symmetric, residual IOAs stem from the asymmetry of the QW  interfaces; for instance, associated to Ga segregation into the AlGaAs layer during the epitaxial growth process. In the case of an asymmetric DQW with QWs with different thicknesses, the AlGaAs layers (that are sources of anisotropies) are not distributed symmetrically at both sides of the tunneling barrier. Thus, the system losses its inversion symmetry  yielding an increase of the RAS strength. The RAS line shapes were compared with reflectance spectra in order to assess the heavy- and light- hole mixing induced by the  symmetry breakdown. The energies of the optical transitions were calculated by numerically solving the one-dimensional Schr\"odinger equation using a finite-differences method. Our results are useful for interpretation of the transitions occurring in both, symmetric and asymmetric DQWs.

\end{abstract}

\maketitle


\section{\label{sec:I} Introduction }
Nanostructures based on double quantum wells (DQWs) constitute an excellent platform to study the interactions between confined energy levels; for instance,  intra-QW transitions between  electrons and holes belonging  to the same QW (direct excitons and trions), and inter-QW transitions occurring for electrons and holes resident in different QW (indirect excitons and indirect trions)  \cite{yuan2018}. The strength of the direct and indirect transitions can be modulated by the application of external perturbations, such as external electric fields. The electric field modifies the tunneling probability leading to the appearance of spatially indirect inter-QW excitons \cite{yuan2018,Miller1984,chen1987} and intra-QW charged excitons (trions) \cite{yuan2018}. Such complex band structures offer  interesting optical phenomena which exhibit polarization dependence.  
In-plane optical anisotropies (IOAs) of AlGaAs/GaAs-based heterostructure have been studied extensively in the past several years by means of reflectance anisotropy spectroscopy (RAS). In the case of single QWs, several sources of IOAs have been investigated: i) absence of 4-fold rotational symmetry axis at the AlGaAs/GaAs interface \cite{krebs1996,chen2002}, ii) asymmetric quantum wells \cite{Koopmans1998}, iii) nonuniform segregation of Ga atoms at AlGaAs/GaAs interfaces \cite{chen2002,braun1997}, iv) application of an  external stress \cite{tang2009,li2019}, and v)  asymmetric barriers \cite{yu2015}. In general, the observed  IOAs strength depends on the QW thickness. For example, for strain-induced IOAS, the RAS response increases with the well thickness \cite{tang2009,li2019}; whereas it decreases with well thickness for  the anisotropies induced exclusively by the interfaces \cite{chen2002,tang2009,yu2015}.

\begin{table*}
	\caption{\label{tab:Tab1} Structure of samples used in this work.}
	\begin{ruledtabular}
		\begin{tabular}{cccc}			
			Sample number      & Well width $ d_{1} (nm)$ & Well width $d_{2} (nm)$ & $\mathrm{Al_{0.15}Ga_{0.85}As}$\\
			\colrule  
			\rule{0pt}{3ex}   
			S1 & 11.87        & 11.87     & Si doped, $\mathrm{n=6\times 10^{18} cm^{-3}}$  \\
		    A1 & 13.85        & 11.87     & Si doped, $\mathrm{n=6\times 10^{18} cm^{-3}}$   \\ 	
			A2 & 23.74        & 11.87     & Si doped, $\mathrm{n=6\times 10^{18} cm^{-3}}$   \\ 	
			A3 & 23.74        & 11.87     & Be doped, $\mathrm{p=5\times 10^{19} cm^{-3}}$   \\ 			
		\end{tabular}
	\end{ruledtabular}
\end{table*}

In DQW-based systems, the relative thickness of the QWs plays an important role in the anisotropy strength; namely, if both QWs have the same thickness (symmetric DQW), the IOA arises from the  contributions  of the  AlGaAs/GaAs(QW)  and GaAs(QW)/AlGaAs interfaces which, in general, are not equivalent. If the QWs thicknesses are different (asymmetric DQW), the system loses its inversion symmetry as a whole along a plane perpendicular to the growth direction an passing through the center of the tunneling barrier; therefore, the interfaces of the DQW system are not distributed symmetrically at both sides of the tunneling barrier, leading to an increase of the IOA signal.

In this paper we report on   RAS experiments on DQWs coupled by a thin barrier (1.98 nm) and different relative thicknesses. We find that by increasing the thickness of one of the QWs relative to the other, we induce a  breakdown of inversion symmetry of the DQW structure which increases the IOA. \\

The rest of the paper  is organized as follows. In Sec.~\ref{sec:II}, a  description of samples and the experimental setup is given. In Sec.~\ref{sec:III}, a description of the numerical model used to calculate the energy of the excitonic transitions of the DQWs and a detailed description of the model used to explain the RAS spectra are discussed. In Sec.~\ref{sec:IV}, we describe the  RAS and photoluminescence (PL) on symmetric and asymmetric DQWs as well as their interpretation on the basis of the breakdown of inversion symmetry. Finally, conclusions are given in Sec.~\ref{sec:V}.

\section{\label{sec:II}Samples AN EXPERIMENTAL SETUP}

\begin{figure}
	\centering
	\includegraphics[width=\columnwidth]{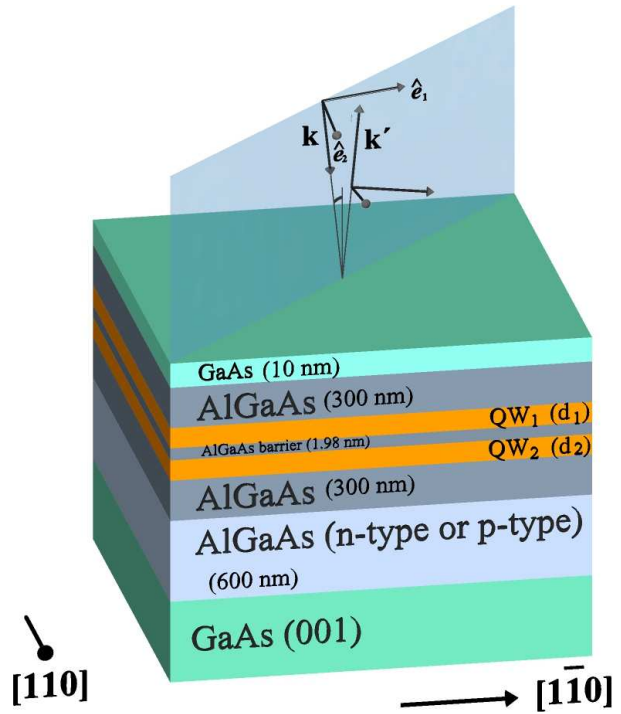}
	\caption{\label{fig:Fig1}  Monochromatic light coming from a tungsten lamp is focused onto the (001) surface of the DQW structures. The angle of incidence is approximately 4 degrees. RAS spectra were obtained by taking the difference between the reflectance for $\hat{e}_{1}$  and $\hat{e}_{2}$  light polarizations. The experiments were performed at 30 K. Thicknesses $d_{1}$ and $d_{2}$ for the samples used in this work are listed in Table~\ref{tab:Tab1}.}
\end{figure}

The samples were grown by molecular beam epitaxy (MBE) on a GaAs (001) semi-insulating substrates covered with a 200 nm-thick smooth GaAs buffer layer  (see Fig.~\ref{fig:Fig1} and Table~\ref{tab:Tab1}). On top of this layer, a 600 nm-thick, n-type (samples S1, A1, A2) or p-type (sample A3)  $\mathrm{Al_{0.15}Ga_{0.85}As}$ layer was grown.  The DQW consists of two undoped GaAs QWs of thicknesses $d_1$ and $d_2$ separated by a 1.98 nm thick $\mathrm{Al_{0.15}Ga_{0.85}As}$ barrier. Sample S1 is a symmetric DQW (i.e., with $d_1 = d_2$) while samples A1, A2 and A3 are asymmetric $(d_1 \neq d_2)$. The whole structures were covered with a 10 nm-thick GaAs capping layer.\\

RAS and reflectance (R) experiments were carried out using for excitation the  monochromatic probe light  obtained by filtering a 70 W tungsten lamp with a monochromator. The monochromatic beam  then passes through a polarizer prism and a photo-elastic modulator in tandem and is  focused onto the sample with a spot size of 5.0 mm of diameter. The sample is chilled at cryogenic temperatures (30 K) by means of a Helium closed cycle cryocooler. The light reflected by the sample is collected and detected by a multialkali photomultiplier tube. A detailed description of the RAS technique can be found elsewhere \cite{LF1993}.

\section{\label{sec:III} Model}
\subsection[Numerical]{Numerical calculations}

The one-dimensional Schr\"odinger equation in the effective mass approach  was solved  using the finite difference method \cite{qwwad,Hamaguchi_2017} to obtain both the transition energies and wave functions of the DQWs. We neglect spin-orbit interaction, which leads to  a $4\!\times\!4$ Luttinger's Hamiltonian \cite{Luttinger1956} with decoupled  heavy- and light- hole states. Taking  into account  the band-edge variation $V(z)$ as well as the dependence of the effective mass $m(z)$ on the composition of the layers  \cite{BenDuke,Hamaguchi_2017,Yutaka1994,LewYanVoon2009}, our system can be described by: 
\begin{equation}\label{eq:Eq1}
\begin{split}
\left[- \dfrac{\hbar^{2}}{2m_{jz}}\dfrac{d^{2}}{dz^{2}}+V(z)\right]\psi_{nj}(z) = E_{nj}\psi_{nj}(z),\\
 j=\text{e,\,hh,\,lh}.
\end{split}
\end{equation}
The Varshni equation was employed for the determination of gap energies of  $\mathrm{Al_{x}Ga_{1-x}As}$  alloys \cite{varshni}.
The conduction and valence band offsets   were chosen as  65\% and 35\%, respectively. Both electron and hole effective masses for GaAs and AlAs  can be found in Refs. \cite{Vurgaftman,molenk,adachi2009},  whereas  for  $\mathrm{Al_{x}Ga_{1-x}As}$  the  Vegard's law was used \cite{donmez}.

\begin{figure}
	\subfigure{%
		\includegraphics[width=\columnwidth]{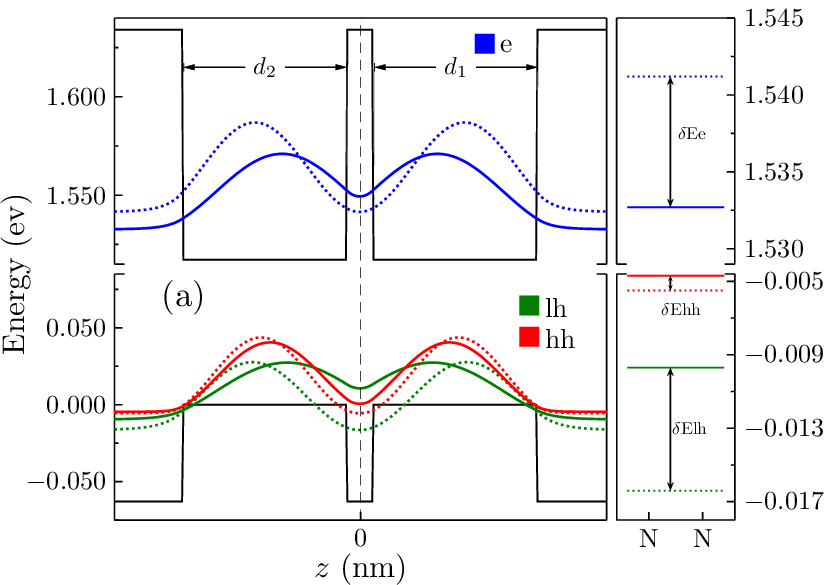}%
		\label{fig:Fig2a}
	}\vspace{-0.25cm}
	\subfigure{%
		\includegraphics[width=\columnwidth]{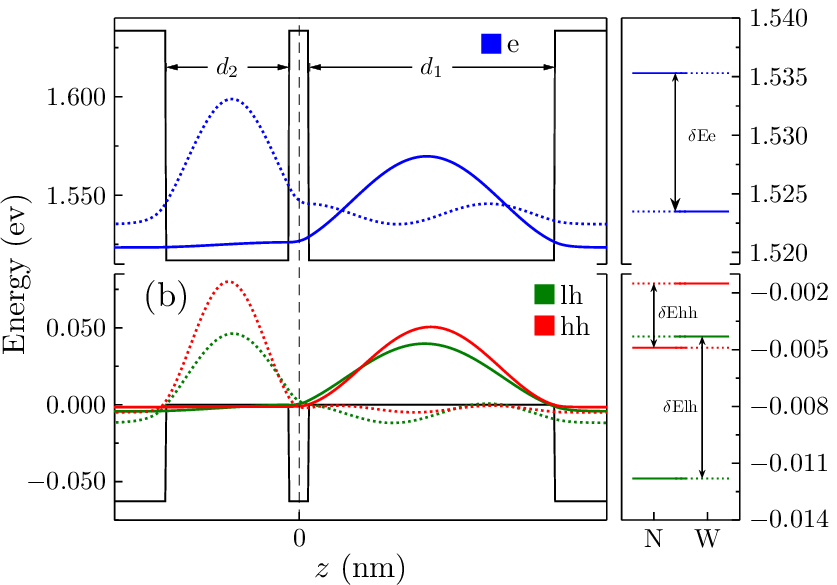}%
		\label{fig:Fig2b}
	}
	\caption{\label{fig:Fig2} Probability density functions (letf panel) and confinment energies (right panel) for (a) sample S1 (symmetric DQW) and (b) Sample A2 (asymmetric DQW). }

\end{figure}

Figure~(\ref{fig:Fig2}) shows the energies and wave functions of the  S1 (Fig.~\ref{fig:Fig2a}) and A1 (Fig.~\ref{fig:Fig2b}) samples calculated by the finite difference model without band nonparabolicity. The transition energies were calculated by taking into  consideration the  exciton binding energy as a function of the  well width \cite{Greene1984,Yutaka1994}. We show that wave functions in asymmetric DQW  are  asymmetrically  distributed  around the tunneling barrier for both  electrons and holes. The barrier width plays an important role in tunneling. For symmetric DQW, as noted above, the electron and heavy- light- holes states are quite similarly localized in the two wells and there is a  splitting of the degenerate single states into doublets. For asymmetric DQW the energies of doublets tend to the singlet state energy of a single QW \cite{SIVALERTPORN2016}.

\begin{table*}
	\caption{\label{tab:Tab2} Comparative of experimental (E) and numerical calculations (N) of transition  energies (in eV). $\mathrm{\delta Ee}$, $\mathrm{\delta Ehh}$ and $\mathrm{\delta Elh}$ corresponds to the difference between electrons, heavy- light holes states, respectively. $\mathrm{\Delta E_n}$ is the numerical calculation of energy splitting for transitions 1 and 2 ($ \mathrm{n}=1,2$).}
	\begin{ruledtabular}
		\begin{tabular}{lp{1.5cm}p{1.5cm} c c c c c c c p{2cm}}			
			Sample &$\quad$e1-hh1 &$\quad$e1-lh1&e2-hh2 &e2-lh2&e3-hh3 &$\mathrm{\delta Ee}$(meV)& $\mathrm{\delta Ehh}$(meV) &  $\mathrm{\delta Elh}$(meV)
			&$\mathrm{\Delta E_n}$(meV) \\
			\colrule  
				\rule{0pt}{3ex} 
			\multirow{2}{*}{S1} &
			(E)1.5328 (N)1.5297&
			(E)1.5380 (N)1.5341& 
			\multirow{2}{*}{(N)1.5394}&   
			\multirow{2}{*}{(N)1.5499}& 
			\multirow{2}{*}{(N)1.5948}& 
			\multirow{2}{*}{8.9}& 
			\multirow{2}{*}{0.9}&
			\multirow{2}{*}{6.9}&
			{$\begin{aligned}[t]
			  \mathrm{\Delta E_1} &=  4.4\\   
			  \mathrm{\Delta E_2} &= 10.5
			\end{aligned}$}\\
			\rule{0pt}{3ex} 
			\multirow{2}{*}{A1} &
			(E)1.5265 (N)1.5273&
			(E)1.5296 (N)1.5314& 
			 \multirow{2}{*}{(N)1.5368}&   
			 \multirow{2}{*}{(N)1.5460}& 
			 \multirow{2}{*}{(N)1.5837}& 
		   \multirow{2}{*}{8.4}& 
            \multirow{2}{*}{1.3}&
		   \multirow{2}{*}{6.4} &
		    {$\begin{aligned}[t] 
		  	\mathrm{\Delta E_1} &= 4.1\\ 
		  	\mathrm{\Delta E_2} &= 9.2 
		  	\end{aligned}$} \\   
		   \rule{0pt}{3ex} 
		   \multirow{2}{*}{A2} &
		   (E)1.5181 (N)1.5190&
		   (E)1.5206 (N)1.5210& 
		   \multirow{2}{*}{(N)1.5330}&   
		   \multirow{2}{*}{(N)1.5394}& 
		   \multirow{2}{*}{(N)1.5460} & 
		   \multirow{2}{*}{11.8} & 
		   \multirow{2}{*}{3.4} &
		   \multirow{2}{*}{7.5 }  &  
		   {$\begin{aligned}[t]
		    \mathrm{\Delta E_1} &= 2.0\\
		   	\mathrm{\Delta E_2} &= 6.4
		   	\end{aligned}$}\\           
		\end{tabular}
	\end{ruledtabular}
\end{table*}

\subsection{\label{subsec:Anisotropy_Model} Anisotropy Model}
GaAs is a cubic crystal of the  symmetry group $T_{d}$. For  a single QW grown on the (001) surface, the $z_{\parallel}(001)$ direction becomes inequivalent to both $x_{\parallel}\left(100\right)$ and $y_{\parallel}\left(010\right)$ directions, hence lowering symmetry from $T_d$ to $D_{2d}$. This group comprises eight operations of symmetry including four that changes
the direction of the $z$-axis. In this symmetry group, the $x$ and $y$ axes remain equivalent and the optical properties are isotropic in the $x - y$ plane. An IOA is expected when the  $z_{\parallel}(001)$ axis loses  its 4-fold rotation symmetry, lowering the symmetry further from  $D_{2d}$ to $C_{2v}$. Different perturbations have been considered to increase the IOA; among them we can count on: uniaxial external stress applied along $\left[110\right]$ direction, ultra-thin layers inserted in one QWs barrier, and external electric fields applied along the $z_{\parallel}\left(001\right)$ direction.
Another mechanism that induces an IOA in a single QW is the non-equivalence of the AlGaAs/GaAs interfaces. As a matter of fact, in a single QW the anisotropies originated at the AlGaAs/GaAs(QW) and at the GaAs(QW)/AlGaAs interfaces are reversed in sign, thus cancelling the IOA ($D_{2d}$ symmetry). However, it is well-known, that, experimentally, they are not exactly equivalent from the structural point of view
\cite{li2019} and a residual IOA is normally observed \cite{krebs1996,chen2002} ($C_{2v}$ symmetry). RAS
 spectra of single QWs have been reported extensively in the literature \cite{chen2002,Koopmans1998,tang2009,li2019,yu2015}.\\

The thickness of the QW plays a fundamental role in the amplitude of the RAS signal. For the anisotropies originated at the interfaces and the ones induced by the segregation of Ga or In (in the case of the inclusion of a thin InAs layer), the RAS amplitude decreases when the thickness of the QW increases \cite{chen2002}. Contrary to this behavior, for the IOA induced by strain, the RAS signal increases with the  thickness of the QW \cite{tang2009}.
In any case, the RAS signal is associated to the mixing between the heavy- and light-hole valence subbands of the QW \cite{Ivchenko1996}. In a perturbative approach the IOA strength is proportional to:

\begin{equation}
\dfrac{
       \langle \psi_{\rm{en}} | \psi_{\rm{hhn}} \rangle
       \langle \psi_{\rm{hhn}}| \mathcal{H} | \psi_{\rm{lhn}} \rangle
       \langle\psi_{\rm{lhn}} | \psi_{\rm{en}} \rangle
      }
      {\rm{\Delta E_{n}}
      },
\label{eq:Eq2}
\end{equation}

where $ \langle \psi_{\rm{en}} | \psi_{\rm{hhn}} \rangle$  and $\langle\psi_{\rm{lhn}} | \psi_{\rm{en}} \rangle$ are the overlap integrals between the  $|\psi_{\rm{en}}\rangle$  electron state in the conduction band and the $|\psi_{\rm{hhn}}\rangle$  and  $|\psi_{\mathrm{lhn}}\rangle$   hole states in the valence band respectively. The  mixing between heavy- and light-hole subbands is $ \langle \psi_{\rm{hhn}}| \mathcal{H} | \psi_{\rm{lhn}} \rangle$  being the  perturbative Hamiltonian. The difference in energy between the hole states before
 their mixing is $\rm{\Delta E_{n} = E_{hhn} - E_{lhn}}$.

As previously mentioned,  $\cal{H}$ could account  for a perturbation  by applied uniaxial stresses, electric fields (built-in or external) and  abrupt or smooth interfaces.
In the case of coupled DQW system, $\cal{H}$ is a  measure of the  asymmetry of the two QWs. 
In a DQW, if both QWs have the same thickness, the electron and hole energy levels are brought
into resonance and their probability density is distributed symmetrically at both sides of the barrier that connects both QWs, as can be seen in Fig.~\ref{fig:Fig2a}. If the AlGaAs/GaAs interfaces of the DQW are equivalent the symmetry of the DQW structure belongs to the $D_{2d}$ crystallographic point group. However, as we mentioned, if the GaAs/AlGaAs interfaces (sources of anisotropies) are non-equivalent, and a residual IOA is induced by the symmetry reduction from $D_{2d}$ to $C_{2v}$ of the whole DQW structure. In this case the number of symmetry operations is reduced to four: the operations that inverts the  $z$-axis, are excluded, thus leading to a mixing of heavy- and light- holes in the valence band, and hence an IOA.
\\

When the thickness of one QWs is larger than the other QW (asymmetric DQW), the energy of the electron and hole states of each QW become different and are no longer in resonance. In this case, the probability density is distributed asymmetrically as can be seen in Fig.~\ref{fig:Fig2b}. Besides of the anisotropic source induced by the GaAs/AlGaAs interfaces as in the symmetric case, for the asymmetric DQW system another mechanism of IOA should be considered. Suppose that the AlGaAs/GaAs interfaces are equivalent. As we mentioned, in this case a symmetric DQW system belongs to the $D_{2d}$ symmetry. If the thickness of one QW is modified, the system changes symmetry from $D_{2d}$ to $C_{2v}$. As can be seen in Fig.~\ref{fig:Fig2b}, the symmetry operations
 that change the sign of the $z$-axis (i.e., a reflection on a
plane perpendicular to $z$), are no longer possible because
the AlGaAs/GaAs interfaces are located at different distances
from the barrier. The number of symmetry operations
of the asymmetric DQW structure is reduced to
the four operations of the symmetry group $C_{2v}$ \cite{Ivchenko2008,glazov2018electron}. In
this case the IOA strength can be modulated by changing
the relative thicknesses of the QWs in the DQW system. This anisotropy is added to the residual IOA produced by the non-equivalence of the interfaces.

\section{\label{sec:IV}Experimental Results}

\begin{figure}
	\centering
	\includegraphics[width=\columnwidth]{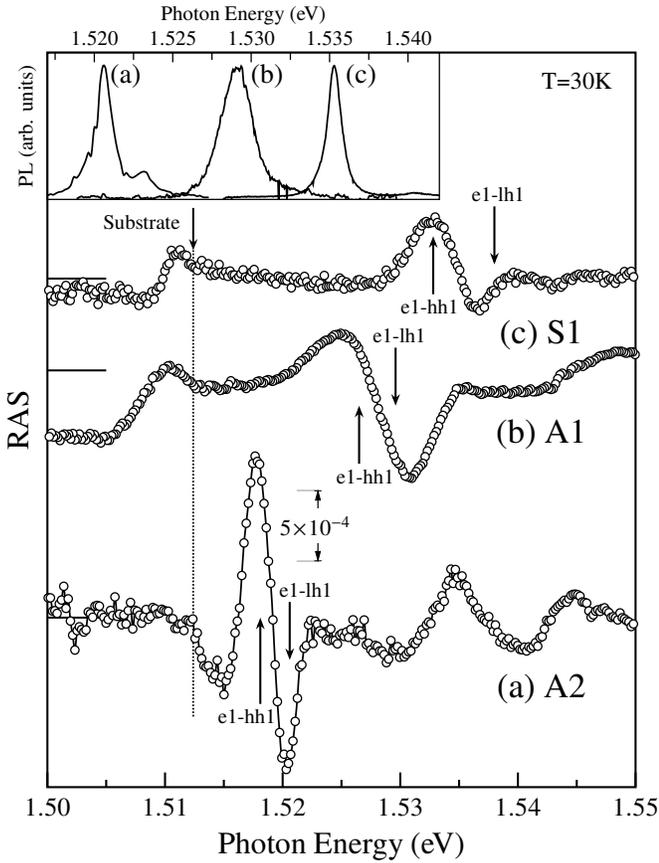}
	\caption{RAS spectra for the (a)-(b) asymmetric and (c) symmetric DQWs. The dashed vertical line indicates the expected energy of the excitonic transition of the GaAs substrate. Above this energy, the optical transitions come from the DQWs. The inset shows the PL spectra measured for each sample. Two peaks can be identified in each spectrum, a larger one associated to the transition e1-hh1 and a much smaller one associated to the e1-lh1 (for spectrum (b) this peak is observed as a shoulder). The energies obtained from the PL spectra are indicated by the arrows in the RAS spectra. Note that the structures associated to e1-hh1 and e1-lh1 increase their strength when the DWQs becomes more asymmetric. The RAS spectra were measured at 30K. } 
	\label{fig:Fig3}
	
\end{figure}

\begin{figure}
	\centering
	\includegraphics[width=\columnwidth]{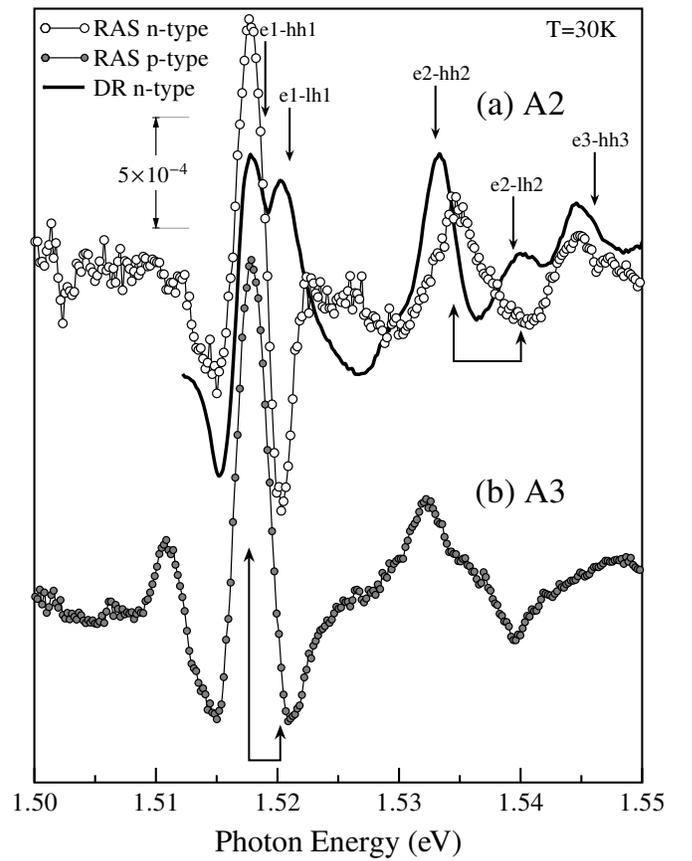}
	\caption{Reflection anisotropy (RAS, open and closed  circles) and differential reflection (DR, solid line) spectra for asymmetric DQWs A2 and A3,  grown on an AlGaAs n-type  and  p-type layer respectively. Note that while for the heavy hole transitions (e1-hh1 and e2-hh2) in the RAS and DR spectra have the same concavity, for light holes transitions (e1-lh1 and e2-lh2) the concavities are opposite and DR spectra shows the highest level transitions. The bottom arrows point to the experimental transitions for the two first levels, whereas the top arrows show the calculated energies to three energy levels.  The RAS and DR spectra were measured at 30K. }
	\label{fig:Fig4}
\end{figure}

Figure~\ref{fig:Fig3} shows the RAS spectra recorded at 30K for the samples S1, A1 and A2 (described in Table~\ref{tab:Tab1}). The  inset plot show the corresponding PL spectra measured at 12K, where the excitonic transitions can be identified by two peaks, the larger one associated with e1-hh1 transition and a much smaller one associated to the e1-hh1. In the case of RAS, the  peaks associated with  these transitions are seen clearer. The e1-hh1 and e1-lh1 transitions determined in the PL spectra are indicated by arrows. These values are in excellent agreement with those determined by the numerical calculations (see Table~\ref{tab:Tab2}). Note that the features associated to each transition have opposite concavities and red-shift when one of the QWs of the system increases its thickness. The evolution of the optical structure is evident: if the DQW structure become asymmetric the corresponding RAS spectra increases its strength.\\

Due to the fact that the wave function probability density of sample $\mathrm{S1}$ is distributed symmetrically along the QWs structure, the IOA strength is expected to be  similar to that obtained for a single QW is expected. In fact, a RAS signal of the order of magnitude of $0.1\times10^{-3}$ has been reported for a 8 nm single QW \cite{chen2002}, which has the same order of magnitude as in spectra of Fig.~\ref{fig:Fig3}(c). This IOA is attributed to the inequivalent AlGaAs/GaAs interfaces along
the DQW structure. \\

As pointed out before, the strength of the IOA signal is produced by an intermixing of the heavy- and light hole states in the valence band which is proportional to $\langle \psi_{\rm{hhn}}| \mathcal{H} | \psi_{\rm{lhn}} \rangle$ according to Eq.~(\ref{eq:Eq2}). For the lowest heavy and light holes levels (n=1), there is an estimated separation in energy of around $\rm{\Delta E_1}$ = 2.0, 4.1 and 4.4 meV for samples A2 (A3), A1 and S1, respectively.  The mixing $\langle\psi_{\rm{hhn}}| \mathcal{H} | \psi_{\rm{lhn}}\rangle$ can be estimated by considering that the transitions are direct (n is the same for the valence and conduction band) and then the overlapping terms in Eq.~(\ref{eq:Eq2}) must be approximately the same for each sample. For transitions n=1 it can be seen in Fig.~\ref{fig:Fig3}, that the amplitude ratio of the spectra between A2 and A1 with respect to S1 are 1.5 and 3.7 respectively. Thus from Eq.~(\ref{eq:Eq2}) we estimate ratios of $\langle\psi_{\rm{hh1}}|{\cal{H}_{\rm{A1}}}|\psi_{\rm{lh1}}\rangle/\langle\psi_{\rm{hh1}}|{\cal{H}_{\rm{S1}}}|\psi_{\rm{lh1}}\rangle\sim 1.4$ for sample A1 and 
$\langle{\psi_{\rm{hh1}}}|{\cal{H}_{\rm{A2}}}|{\psi_{\rm{lh1}}}\rangle/\langle{\psi_{\rm{hh1}}}|{\cal{H}_{\rm{S1}}}|{\psi_{\rm{lh1}}}\rangle\sim 1.7$ for sample A2.

In order to elucidate the physical origin of the RAS in the asymmetric DQWs, we compare in Fig.~\ref{fig:Fig4} the RAS and the DR spectra of the asymmetric sample A2. DR spectra is obtained by the numerical subtraction of the reflection (R) spectra recorded at 30K and 300K followed by the normalization to the 300K spectrum. The subtraction highlights the excitonic features, which are very weak (and energy shifted) at 300K. The comparison between RAS and DR spectra allows us to contrast the contribution of the heavy and light holes transitions. Around 1.5175 eV, the DR spectrum shows two peaks corresponding to e1-hh1 and e1-lh1 transitions. Note that in the RAS spectrum, the structure associated to e1-hh1 transition has the same concavity as the corresponding for the DR spectrum, while the e1-lh1 transition has the opposite concavity. This  is an indication of the transfer of oscillator strength between the levels due to the intermixing of heavy- and light- holes, thus supporting our anisotropy model. The same behavior applies for the e2-hh2 and e2-lh2 transitions at around 1.5375 eV. Transition e2-hh3 is also indicated and it has the same concavity for RAS and DR spectra, as in the case of the e1-hh1 and e2-hh2 transitions.

The arrows at the bottom of Fig.~\ref{fig:Fig4} indicate the energy of the states en-hhn and en-lhn (for n =1 and 2) obtained from the  maximum and minimum of the RAS spectrum. 
From the numerical calculation results summarized in  Table~\ref{tab:Tab2}, the energy splitting between transitions  en-hhn and  en-lhn  are $\rm{\Delta E_1}=2.0$ meV and $\rm{\Delta E_2}=6.4$ meV. In accordance with Eq.~(\ref{eq:Eq2}), the IOA amplitude is proportional to $1/\rm{\Delta E_n}$. Considering the same valued for the overlapping and the mixing $\langle \psi_{\rm{hhn}}| \mathcal{H} | \psi_{\rm{lhn}} \rangle$ we estimate an amplitude ratio of 3.25 between these transitions. This value is close to the value of 3.9 obtained by the RAS spectrum of Fig.~\ref{fig:Fig4} supporting our interpretation.

Finally, we discuss the possible contribution to the RAS amplitude by an built-in electric field accross the DQW. To study this contribution, we have compared the RAS spectra of asymmetric samples A2 and A3. The difference between them is the doping of the AlGaAs layer (see Fig.~\ref{fig:Fig1}). While for A2 it is n-type, for A3 it is p-type. Assuming that the built-in electric field  originates from charge transfer between surface states and  the AlGaAs doped layer (n or p), this field is expected to have opposite signs for samples A2 and A3. Thus, the linear contribution of the electric field to the RAS should be reversed in sign for such samples. Figure~\ref{fig:Fig4} shows the comparison between the RAS spectra of samples A2 and A3 obtained from the difference in reflectivity between the  $\left[1\overline{1}0\right]$ and $\left[110\right]$ crystallographic directions. As can be seen, RAS spectra are equivalent  in shape and have the same sign, thus indicating that the contribution of the electric field to the RAS signal is very small. We conclude, thus, that the dominant contribution to RAS spectra is the asymmetry of the DQW system.\\

\section{\label{sec:V} Conclusion}
We presented in this work a detailed study of the evolution with QW thickness of the RAS spectra of a coupled DQW system. The RAS signal increases its strength with the degree of asymmetry of the QWs that constitutes the DQW system. This behavior is attributed to the reduction of the symmetry of the electronic states in the DQWs with QWs of different thicknesses. The nature of the transitions was identified by using PL spectroscopy and numerical calculations. We believe that the result presented in this work are important and will be  useful in  understanding of the evolution of the optical transition induced by the breakdown of translation symmetry in  asymmetric DQWs.

\section{ACKNOWLEDGMENTS}
We would like to thank L. E. Guevara-Mac\'ias, E. Ontiveros, F. Ram\'irez-Jacobo and J. Gonzalez-Fortuna for
their skillful technical support. This work was supported by Consejo Nacional de Ciencia y Tecnolog\'ia (grants
FC-2093-2018, CB-256578-2016, CB-30406-2018, CB-252867 and 299552).

%

\end{document}